\documentclass{article}

\usepackage{amssymb}
\usepackage{ulem}
\usepackage{xcolor}
\usepackage{arxiv}
\usepackage{newtxtext}
\usepackage[utf8]{inputenc} 
\usepackage[T1]{fontenc}    
\usepackage{url}            
\usepackage{booktabs}       
\usepackage{amsfonts}       
\usepackage{nicefrac}       
\usepackage{microtype}      
\usepackage{lipsum}
\usepackage{natbib}
\usepackage{amsmath}
\usepackage{tikz}
\usetikzlibrary{bayesnet}
\usepackage{graphicx}
\usepackage{xr-hyper}
\usepackage[draft]{hyperref}
\usepackage{tabularx}
\newcolumntype{C}{>{\centering\arraybackslash}X}
\usepackage{makecell}
\usepackage{caption}
\usepackage{xspace}
\usepackage{fancyvrb}
\usepackage{hyperref}
\usepackage{bibunits}
\usepackage{lineno}
\newcommand{\labelms}[1]{}

\newcommand{\refms}[1]{}

\graphicspath{ {./images/} }

\title{\textit{NExON-Bayes}: A Bayesian approach to network estimation informed by ordinal covariates}

\author{
 Joseph Feest \\
  MRC Biostatistics Unit\\
  University of Cambridge\\
  Cambridge, CB2 0SR, UK \\
  \texttt{joseph.feest@mrc-bsu.cam.ac.uk} \\
   \And
 Hélène Ruffieux \\
  MRC Biostatistics Unit\\
  University of Cambridge\\
  Cambridge, CB2 0SR, UK \\
  \texttt{helene.ruffieux@mrc-bsu.cam.ac.uk} \\
  \And
 Camilla Lingj\ae rde  \\
 Department of Mathematics \\
  University of Oslo\\
  Oslo, N-0316, Norway\\
  \texttt{camiling@math.uio.no} \\
  \And
 Xiaoyue Xi \\
  MRC Biostatistics Unit\\
  University of Cambridge\\
  Cambridge, CB2 0SR, UK \\
  \texttt{xiaoyue.xi@mrc-bsu.cam.ac.uk} \\
}
\begin{document}
\maketitle

\abstract{\textbf{Motivation:} In heterogeneous disease settings, accounting for intrinsic sample variability is crucial for obtaining reliable and interpretable omic network estimates. However, 
most graphical model analyses of biomedical data assume homogeneous conditional dependence structures, potentially leading to misleading conclusions. To address this, we propose a joint Gaussian graphical model that leverages sample-level ordinal covariates (e.g., disease stage) to account for heterogeneity and improve the estimation of partial correlation structures.
\\
\textbf{Results:} Our modelling framework, called NExON-Bayes, extends the graphical spike-and-slab framework to account for ordinal covariates, jointly estimating their relevance to the graph structure and leveraging them to improve the accuracy of network estimation. To scale to high-dimensional omic settings, we develop an efficient variational inference algorithm tailored to our model. Through simulations, we demonstrate that our method outperforms the vanilla graphical spike-and-slab (with no covariate information), as well as other state-of-the-art network approaches which exploit covariate information. 
Applying our method to reverse phase protein array data from patients diagnosed with stage I, II or III breast carcinoma, we estimate the behaviour of proteomic networks as cancer progresses. Our model provides insights not only through inspection of the estimated proteomic networks, but also of the estimated ordinal covariate dependencies of key groups of proteins within those networks, offering a comprehensive understanding of how biological pathways shift across disease stages. \label{ref:some_string}
\\
\textbf{Availability and Implementation:} A user-friendly R package for NExON-Bayes with tutorials is available on Github at \url{github.com/jf687/NExON}.}
\section{Introduction}\label{Introduction}
Graphical models are useful tools to uncover complex dependence structures between molecular variables and shed light on the biological mechanisms underlying disease. In Gaussian graphical models (GGMs), typically employed for the analysis of omic data, the partial correlations between nodes can be derived from the off-diagonal elements of the inverse covariance (precision) matrix. Various approaches for GGM estimation have been developed using frequentist and Bayesian approaches. Frequentist approaches are based on penalised maximum likelihood estimation and include neighbourhood selection \citep{Meinshausen06Lasso}, the graphical lasso \citep{friedman2008sparse} and the graphical SCAD \citep{Fan2009}.
Bayesian methods place a shrinkage prior on the precision matrix entries and include the Bayesian graphical lasso \citep{wang2012bayesian}, the graphical horseshoe \citep{Li2019} and the graphical spike-and-slab \citep{wang2015scaling}. These methods all assume homogeneous dependence structures across all samples, an assumption that is often unrealistic in practical settings where, for example, samples might be collected from patients across different conditions, such as different disease severities or treatment regimes. 

More recently, methods have been developed to jointly estimate graphical models across multiple networks, reflecting such conditions. Joint methods estimate network structures through sharing information between networks when appropriate, while ideally retaining network-specific differences. 
Several approaches for joint estimation of multiple GGMs have been proposed. Frequentist approaches extend the penalised likelihood framework to add penalty parameters that encourage similarity between network-specific precision matrices \citep{guo2011joint,ma2016joint,lingjaerde2023stabjgl}, whilst Bayesian approaches use priors to encourage the potential sharing of network structures \labelms{2.1}[\citealp{peterson2015bayesian}, \citealp{Li2019, lingjaerde2024scalable}]. Such methods have been used, for example, to estimate the proteomic networks of breast cancer patients with different subtype diagnoses, where the expected similarities between networks are leveraged in the joint estimation \citep{lingjaerde2023stabjgl}. 

Existing joint estimation approaches often treat networks as exchangeable, which in some cases can inhibit statistical power. For example, when the covariates used to define subgroups have a natural ordering, the assumption of exchangeability both omits useful information that could be leveraged in the estimation and limits the potential insight into how those covariates affect graphical structure. 
A related set of more recent approaches build on this idea by making GGMs dependent on covariates. Despite being applicable in a plethora of settings, these approaches do not ensure the positive definiteness of the precision matrix and often lack scalability, which seems to hamper widespread adoption \labelms{s2.6}{\citep{ni2022bayesian, niu2024covariate}.

Here we present NExON-Bayes, a Bayesian approach to network estimation informed by ordinal covariates. \labelms{s1.1b}NExON-Bayes is thus applicable to settings with explicit ordinal covariates or where an ordinal structure can reasonably assumed. These covariates encode auxiliary, non-omic data, which can represent a varied information, such as tumour progression with a covariate taking values in $\{1,2,3\}$ corresponding to stage I, II \& III cancers, 
therapeutic response categories with covariate taking values in 
$\{1,2,3,4\}$ corresponding to the response category \citep{eisenhauer2009new}, 
or genotype data with covariate taking values in $\{0,1,2\}$ corresponding to the number of copies of a specified allele.\labelms{e1.1b}
NExON-Bayes builds on the standard graphical spike-and-slab model of \citet{wang2015scaling}, and introduces a sub-model that characterises the dependence between edge inclusion probabilities and sample-level ordinal covariate data, rather than modelling the entries of the precision matrix, as done in other covariate-dependent approaches.
This modelling framework allows us to develop a deterministic inference algorithm that is more scalable than existing alternatives in the literature.

This paper is organised as follows. 
Section 2 (Methods) introduces the terminology and notation used throughout the article, our proposed model, and the variational inference algorithm. Next, Section 3 (Results) assesses the performance of our approach by benchmarking the model against several competitors through simulations, and investigates its benefits in a real-life setting where we estimate the proteomic networks of breast carcinoma patients. Finally, Section 4 (Discussion) summarises our contributions and avenues for extensions of this work.
\section{Methods}\label{Methods}
\subsection{Terminology and notation}\label{notation}
A graphical model is a probabilistic framework that represents dependencies among a set of random variables. It is represented by $\mathcal{G = (V},E)$, where $\mathcal{V} = \{1, \ldots, P\}$ is the set of nodes (variables) and $E$ is the set of undirected edges between them.
In a Gaussian graphical model (GGM) the nodes represent random variables from a multivariate Gaussian distribution, whilst the edges represent conditional dependence between the variables that they connect. A GGM assumes that 
$N$ samples of $P$ variables are independently and identically distributed (i.i.d.) according to 
\begin{align}
    {\boldsymbol{y}_n} \sim \mathcal{N}_P(\boldsymbol{0}, \boldsymbol{\Omega}^{-1} ), \quad n = 1,...,N,\nonumber
\end{align}
where $\boldsymbol{\Omega}$ is the positive definite inverse covariance matrix, known as the \textit{precision matrix}. 
In this work, we consider settings in which each of the $N$ samples is characterised by a sample-level ordinal covariate $a$ taking values in a set $\mathcal{A}$.  \labelms{s1.1c}To leverage this information, we assume sample homogeneity only within groups defined by equal covariate value, rather than across all samples.
Specifically,  
\begin{align}
    {\boldsymbol{y}_n^{(a)}} \sim \mathcal{N}_P(\boldsymbol{0}, {(\boldsymbol{\Omega}^{(a)})^{-1}} ), \quad n = 1,..., N_a,\label{eq:normal_dist}
\end{align}
where $\boldsymbol{\Omega}^{(a)}$ is the precision matrix specific to the $N_a$ samples with covariate value $a \in \mathcal{A}$.\footnote{When referring to the covariate value, we use `$a$', and when referring to networks or data that correspond to a covariate value $a$ we use `$(a)$'.} \labelms{e1.1c} A covariate-specific dataset can then be denoted as $\boldsymbol{Y}^{(a)} = (\boldsymbol{y}^{(a)}_1, ..., \boldsymbol{y}^{(a)}_{N_a})$.

Partial correlations between variables can be calculated from these precision matrices, which measure the conditional dependence between variable pairs. Specifically, partial correlation quantifies the correlation between a pair of variables while conditioning on all other variables in the system, thereby isolating direct effects by accounting for potential confounding variables, mediators or colliders. Mathematically, the partial correlation between nodes $i$ and $j$ in network $(a)$ is given by

\begin{equation}
    \rho^{(a)}_{ij} = -\frac{\boldsymbol{\Omega}^{(a)}_{ij}}{\sqrt{\boldsymbol{\Omega}^{(a)}_{ii} \, \boldsymbol{\Omega}^{(a)}_{jj}}},\quad i \neq j, \nonumber
\end{equation}
where $\boldsymbol{\Omega}^{(a)}_{ij}$ is the $ij$'th entry of $\boldsymbol{\Omega}^{(a)}$.
Thus, if an entry of the precision matrix is non-zero, there is conditional dependence between the corresponding nodes (and vice versa). The edge set $E^{(a)}$ of the corresponding GGM is then given by $E^{(a)} = \{\, (i, j) \mid \boldsymbol{\Omega}^{(a)}_{i j} \neq 0 \,\}$. Note that because $\boldsymbol{\Omega}^{(a)}$ is symmetrical, $(i,j) \in E^{(a)} \Leftrightarrow (j,i) \in E^{(a)}$.

\subsection{Graphical spike-and-slab model}\label{GSSL}
\begin{figure}[t!]
  \centering
\begin{tikzpicture}
\node[obs] (yn) {$\boldsymbol{y}_n^{(a)}$} ; 
\node[latent, right = of yn] (omegaij) {$\mathbf{\Omega}_{ij}^{(a)}$} ; 
\node[latent, above = of yn] (omegaii) {$\mathbf{\Omega}_{ii}^{(a)}$} ;
\node[latent, right = of omegaij] (deltaij) {$\delta_{ij}^{(a)}$} ;
\node[latent, right = of deltaij] (betaij) {$\beta_{ij}$} ;
\node[latent, below = of deltaij, yshift = -0.4cm] (zetaij) {$\zeta_{ij}$} ;
\node[obs, above = of deltaij] (a) {$a$} ;
\node[latent, right = of betaij] (sigma2) {$\sigma^2$};

\edge {omegaij,omegaii} {yn} ; 
\edge {deltaij} {omegaij} ; 
\edge {a, betaij, zetaij}{deltaij} ; 
\edge {sigma2} {betaij} ; 

\plate[inner sep=0.15cm]{plate1} {(yn)} {$n = 1, \ldots, N_a$} ;
\plate[inner sep=0.25cm]{plate2} {(omegaij) (deltaij) (zetaij) (betaij)} {$i,j = 1,\ldots,P,\ i<j$} ;
\plate[inner sep=0.15cm]{plate3} {(omegaii)} {$i = 1, \ldots, P$} ;
\plate[inner sep=0.35cm,
    label={[label distance=-7mm]north:$a \in \mathcal{A}$}
] {plate4} {(plate1) (plate3) (a) } {};
\end{tikzpicture}

\caption{Schematic representation of the NExON-Bayes model. Shaded nodes are observed, and non-shaded are latent variables that are inferred.}
\label{fig:graphmethod}
\end{figure}

To induce sparsity, our proposed approach, NExON-Bayes, relies on a continuous spike-and-slab prior on the off-diagonal elements of each $\mathbf{\Omega}^{(a)}$, similarly as in \cite{wang2015scaling}. This is achieved by using a mixture of two Gaussian distributions, one with very low variance (the spike) and the other with high variance (the slab).
A binary edge inclusion indicator is also introduced within this prior, which is denoted as $\delta_{ij}^{(a)}$, corresponding to the $ij$'th entry of $\mathbf{\Omega}^{(a)}$. Edge inclusion is indicated by $\delta_{ij}^{(a)} = 1$, while edge exclusion is indicated by $\delta_{ij}^{(a)} = 0$. The spike-and-slab prior is thus represented as
\begin{equation}
\boldsymbol{\Omega}_{ij}^{(a)}  \mid \delta_{ij}^{(a)}  \sim  \delta_{ij}^{(a)} \mathcal{N}(0, \nu_1^2) + \big(1 - \delta_{ij}^{(a)} \big) \mathcal{N}(0, \nu_0^2), \quad \nu_0^2 \ll \nu_1^2, \nonumber
\end{equation}
for  $1 \leq i < j \leq P$. A symmetric matrix is formed by mirroring this lower triangle across the diagonal. \labelms{s1.11}Note that a \textit{mixture-of-Gaussians} spike-and-slab prior is used, following \citet{wang2015scaling}, rather than the alternative option of a point-mass spike. This choice preserves positive definiteness in the precision matrix during iterative inference through conjugate updating \citep[see Supplementary Material  1.1 \& 1.1.1;][]{george1997approaches}.\labelms{e1.11}

\subsection{Dependence on ordinal covariates}\label{Dooc}
NExON-Bayes introduces dependence on an ordinal covariate via a probit submodel placed on the edge inclusion indicator which creates covariate dependence on an edge's probability of inclusion:
\begin{equation}
        \delta_{ij}^{(a)} \mid \zeta_{ij} ,\beta_{ij}\sim \text{Bernoulli}\left\{\Phi(\zeta_{ij} +  a \beta_{ij})\right\},\label{eq:delt_re}\nonumber
\end{equation}
where $\Phi(\cdot)$ is the standard normal cumulative distribution function \labelms{1.1a}and $a$ is the ordinal covariate. The parameter $\beta_{ij}$ is an edge-specific regression coefficient quantifying the influence of the covariate on the inclusion of edge $(i,j)$. Specifically, $\beta_{ij} = 0$ means no relationship between covariate value and edge inclusion probability, i.e., edge $(i,j)$ has the same probability of being present or absent in \textit{all} networks. A positive value of $\beta_{ij}$ means that edge inclusion is more likely in networks with high covariate values and vice versa. 
In addition, $\zeta_{ij}$
captures the varying baseline inclusion probability of edges and acts as an offset within the probit submodel. We assign a Gaussian prior distribution to $\beta_{ij}$,
\begin{equation*}
    \beta_{ij} \sim \mathcal{N}(0, \sigma^2),
\end{equation*}
where its hyperparameter $\sigma^{-2}$ is assigned a Gamma prior,
\begin{align}
    \sigma^{-2} &\sim \text{Gamma}(\alpha_\sigma, \beta_\sigma), \quad \alpha_\sigma = \beta_\sigma = 2.\nonumber
\end{align}
\labelms{s1.2a}
Whilst a full multivariate normal could be maintained for all $\beta$ variables, this mean-field approximation reduces the computational complexity of the algorithm, aiding scalability to high-dimensional settings by reducing runtime and memory requirements. Moreover, the model's hierarchical structure implicitly preserves some dependency via the precision matrices (see Figure \ref{fig:beta_mat} and analysis in Section \ref{BorInfo} (Results)). 
\labelms{e1.2a}
Each $\zeta_{ij}$ is also assigned a Gaussian prior,
\begin{align}
    \zeta_{ij} \sim \mathcal{N}(n_0, t_0^2),\nonumber
\end{align}
where the parameters $n_0$ and $t_0^2$ are fixed following \cite{10.1093/biostatistics/kxae021}
to prior beliefs about the mean and variance of the number of edges in the networks, and thus the overall sparsity. The hierarchical structure of these priors is presented schematically in Figure \ref{fig:graphmethod}.

\subsection{Variational inference algorithm}\label{VI}

Stochastic sampling algorithms like Markov chain Monte Carlo (MCMC)
are typically computationally demanding for large graphical models \citep{wang2015scaling}. Instead, we employ deterministic inference algorithms which have been successfully applied in graphical modelling contexts, see, e.g.,  \citet{li2019expectationconditionalmaximizationapproach}, \citet{lingjaerde2024scalable},  \citet{10.1093/biostatistics/kxae021}.
Specifically, we develop a variational Bayes expectation conditional maximisation (VBECM) algorithm.
The true posterior distribution can be written as $p(\underline{\boldsymbol{\Omega}}, \boldsymbol{\Theta}\mid \boldsymbol{y})$, where $\underline{\boldsymbol{\Omega}} = \{\boldsymbol{\Omega}^{(a)}\}_{a \in \mathcal{A}}$ and $\boldsymbol{\Theta}$ is the set of all variables outside of the precision matrix. This true distribution is approximated by  using a variational distribution with mean-field approximation:
$$
q (\underline{\boldsymbol{\Omega}}, \boldsymbol{\Theta}) \label{eq:v_dist}  = \prod_{a \in \mathcal{A}} q(\boldsymbol{\Omega}^{(a)}) \prod_{a\in \mathcal{A}}\prod_{i<j} q(\delta^{(a)}_{ij}, z_{ij}^{(a)})\prod_{i<j} q(\zeta_{ij}) q(\sigma^{-2}) \prod_{i<j} q(\beta_{ij}),\nonumber
$$
where   
$z_{ij}^{(a)}$ is introduced by reparametrising $\delta_{ij}^{(a)}$ as:
\begin{align*}
    \delta_{ij}^{(a)} \mid& z_{ij}^{(a)} = \mathbb{I}\{z_{ij}^{(a)} > 0 \},\\
    \text{where} \quad z_{ij}^{(a)} \mid& \zeta_{ij}, \beta_{ij} \sim \mathcal{N}(\zeta_{ij} + a\beta_{ij} , 1).
\end{align*}
The inference problem can then be reframed as finding an approximation to the true distribution, denoted  $q(\underline{\boldsymbol{\Omega}}, \boldsymbol{\Theta})$, 
that is as close as possible to the true posterior distribution by minimising the reverse Kullback--Leibler divergence. 
This is equivalent to maximising the 
\textit{evidence lower bound} (ELBO),
\begin{equation}
     \mathcal{L}(q) = \mathbb{E}_{q(\underline{\boldsymbol{\Omega}}, \boldsymbol{\Theta})} \{\log p(\boldsymbol{Y}, \underline{\boldsymbol{\Omega}}, \boldsymbol{\Theta})\} 
     - \mathbb{E}_{q(\underline{\boldsymbol{\Omega}}, \boldsymbol{\Theta})} \{\log q(\underline{\boldsymbol{\Omega}}, \boldsymbol{\Theta})\}\nonumber
\end{equation}
\citep{bishop2006pattern}.
The optimal distribution for the $n$-th factor of $\boldsymbol{\Theta}$, denoted $q_n(\boldsymbol{\Theta}_n)$ is found by iteratively taking the expectation of the logarithm of the joint distribution over all the other factors,
\begin{equation}
    \log q^*_n(\boldsymbol{\Theta}_n) = \mathbb{E}_{q(\underline{\boldsymbol{\Omega}}, \boldsymbol{\Theta} \backslash \boldsymbol{\Theta}_n )} \{\log p(\boldsymbol{Y, \underline{\Omega}, \Theta})\} + \text{constant},\nonumber
\end{equation} 
where the $\text{constant}$ absorbs all terms not depending on $\boldsymbol{\Theta}_n$.
\labelms{s1.3}Inferring the precision matrices requires more attention, as the variational distribution of each $\boldsymbol{\Omega}^{(a)}$ is intractable and therefore necessitates a conditional-maximisation step where each $\boldsymbol{\Omega}^{(a)}$ is updated in a blockwise fashion according to 
\cite{wang2015scaling}. 
Importantly, this updating scheme ensures that positive definiteness is maintained at each iteration, as detailed in Supplementary Material 1.1.1. The update rules for each variable, along with their derivations and ELBOs are given in supplementary Material 1.1 and 1.2.\labelms{e1.3}

Finally, we use the extended Bayesian Information Criteria ($\text{BIC}_\gamma$) to select the spike variance $\nu_0$. $\text{BIC}_\gamma$
extends the Bayesian Information Criteria (BIC) with an additional regularisation term
. The $\text{BIC}_\gamma$ of a network-specific graphical estimation, $\boldsymbol{\Omega}^{(a)}_{est.}$, can be found using the following equation: 
\begin{equation}
\text{BIC}_\gamma(\boldsymbol{\Omega}^{(a)}_{est.}) = \underbrace{-2 \ell(\boldsymbol{\Omega}^{(a)}_{est.}) + |E^{(a)}|\log(N_a)}_{\text{=BIC}} + 4 \gamma |E^{(a)}|\log(P),\nonumber
\end{equation}
where $\ell(\cdot)$ is the log likelihood  and $|E^{(a)}|$ is the number of edges present in estimated network $(a) $ \citep{foygel2010extended}. \labelms{s1.4}The regularisation parameter $\gamma$ controls the sparsity of the estimates. \citet{chen2008extended} have shown that, provided that \( 0 \leq \gamma \leq 1 \), the asymptotic properties of \( \text{BIC}_\gamma \) are maintained and valid model selection is achieved; we follow the default setting employed by these authors, 
namely \( \gamma = 0.5 \), which balances model complexity and selection accuracy in high-dimensional contexts \citep{foygel2010extended}.
\labelms{e1.4}
To optimise the values of $\nu_0$, we use line searches to find a value of $\nu_0^{(a)}$ that minimises $\text{BIC}_\gamma(\boldsymbol{\Omega}_{est.}^{{(a)}})$ for each network $(a)$ in the vanilla graphical spike-and-slab model, which gives a strong estimate for expected sparsity.

\section{Results}\label{Results}

\subsection{Simulated Data}\label{SimData}

We use simulation studies to assess the performance of NExON-Bayes by analyzing how well it can recover the true precision matrix from which simulated data is generated. In the main scenario that we consider, linear relations are simulated between probit edge inclusion probabilities and ordinal covariates taking values $a\in \mathcal{A} = \{1,2,3, 4\}$. We correspondingly generate four precision matrices, each with $P = 100$ nodes, \labelms{1.6}from which we sample normally distributed data according to Equation~\eqref{eq:normal_dist}, where the number of observations for each dataset is $N_a = 150$. To start, network $(1)$ is simulated as scale-free, with $100$ edges ($|E| = P$) and thus a sparsity of $0.02$. `Scale-free' refers to a graph that has a degree distribution that follows a power law, which leads to few nodes having several connected edges (hubs) and most nodes having very few connected edges. This trait emulates generic biological networks, which are often scale-free \citep{barabasi2003scale}. The other networks are then created by altering the entries of this first network. Specifically, the zero-valued partial correlations corresponding to $50$ edges that are not present in network $( 1)$ are simulated to linearly increase with the value of the covariate $a$. Similarly, the non-zero valued partial correlations corresponding to $50$ edges that are present in network $(1)$ are simulated to linearly diminish to zero, such that they are not present ($\rho_{ij} = 0$) in network $(4)$. The remaining $50$ edges present in network $(1)$ are present in all four networks, with unchanged partial correlations. We call edges with increasing partial correlation `appearing' edges and those with decreasing partial correlation `disappearing' edges. 

Simulating precision matrices in this way gives rise to an interesting artefact: appearing and disappearing edges have smaller precision matrix entries in networks $(2)$ and $(3)$ and 
are thus more difficult to uncover. 
A more detailed description of how these networks are generated can be found in Supplementary Material 2.1 and seen schematically in Supplementary Figure S1.

\subsection{Borrowing information across networks using joint modelling}\label{BorInfo}

 \begin{figure}
     \centering
     \includegraphics[width=0.35\paperwidth]{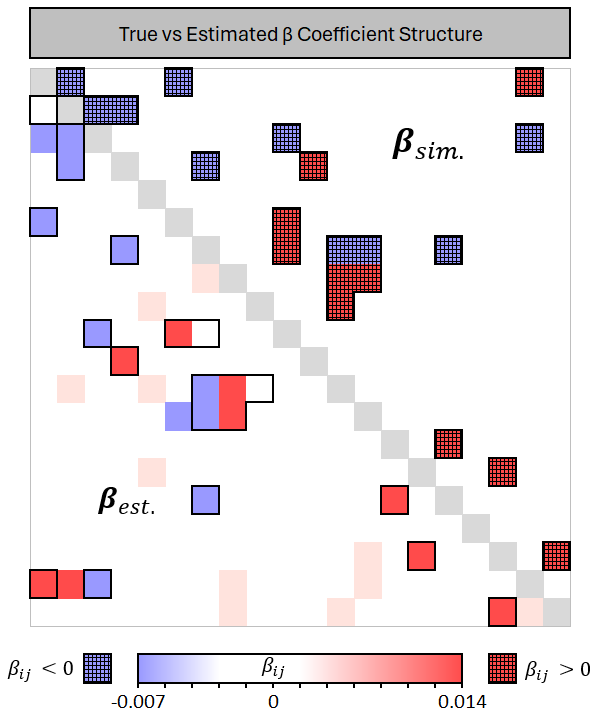} 
     \caption{Posterior mean estimates of each $\beta_{ij}$ (lower diagonal part) versus simulated coefficients of the ordinal covariate   (upper diagonal part). Note that there is no real `true' $\mathbf{\beta}$, as the precision matrices are not simulated from the model. However, there is a known structure based on positive ($\beta_{ij}> 0$) and negative ($\beta_{ij}< 0$) dependencies between covariate and partial correlation. The outline of the true structure is overlain on the estimated structure to ensure a clear comparison.}
     \label{fig:beta_mat}
 \end{figure}
 
We now use this simulated data to compare the performance of NExON-Bayes to the standard single-network Bayesian graphical spike-and-slab model (SSL) \citep{10.1093/biostatistics/kxae021,li2019expectationconditionalmaximizationapproach}, upon which our extension is built. Specifically, we evaluate whether introducing a direct relationship between a covariate and the posterior probability of edge inclusion (PPI) indeed boosts statistical power -- as a first sanity check that the model behaves as it is meant to. 

Table \ref{tab:AUCs} indicates improved area under the curve (AUCs) with NExON-Bayes compared to SSL applied to each network individually. This holds for the estimates of all four networks, with a marked difference in the second and third networks -- suggesting that the borrowing of information enabled by the joint modelling is particularly beneficial when NExON-Bayes is used for estimating networks with weaker signals.

\begin{table}[!ht]
    \centering
    \caption{AUCs for NExON-Bayes and SSL for each simulated network. Standard deviations are in brackets, and the highest AUC for each network is shown in bold. }

    \begin{tabular}{ccc}
    \hline
         Network & SSL & NExON-Bayes  \\ \hline
        1 & 0.966 (0.004) & \textbf{0.974} (0.003)  \\
        2 & 0.870 (0.007) & \textbf{0.967} (0.003)  \\
        3 & 0.857 (0.009) &\textbf{0.946} (0.006)  \\ 
        4 & 0.980 (0.003) & \textbf{0.993} (0.001)  \\ \hline
    \end{tabular}
    \label{tab:AUCs}

\end{table}

The level of information sharing is controlled by the edge-specific regression coefficient $\beta_{ij}$. Not only is the estimation of $\beta_{ij}$ useful in increasing statistical power, it also provides insight on sample-level heterogeneity in terms of the effect of the sample-level covariate on the graphical structure. 
We now consider a smaller simulated setting (with $P = 20$) focused on assessing the ability of NExON-Bayes to recover these edge-specific regression coefficients.
The results shown in Figure \ref{fig:beta_mat} indicate that, 
\labelms{s1.2b} 
while the model does make a mean-field assumption of each $\beta_{ij}$, NExON-Bayes successfully captures the changes of edges associated with covariates, information which, as we will illustrate in the cancer data analysis (see \textit{Real-World Application}), can enable clinically meaningful interpretation of the estimated network structures. Such insights are not captured intrinsically by state-of-the-art single or joint network modelling approaches.

\subsection{Comparison to existing joint and covariate-dependent approaches}\label{Comp}

The performance of NExON-Bayes is then compared against three  other models that are representative of, or similar to, the methodology incorporated into our proposal. The first is the \textit{covariate dependent graphical estimation} model (covdepGE) of \cite{helwig2024algorithm}, which models the precision matrix as a function of the covariate. This approach models the dependence on the covariate via the entries of the precision matrix directly, rather than via the PPI of each edge (as is done in NExON-Bayes). The second approach is the \textit{Bayesian Joint Spike-and-Slab Graphical Lasso} (SSJGL) of \cite{Li2019}, which treats the networks as exchangeable, thereby ignoring any inherent ordering among them. Both of these models are extensions of the graphical spike-and-slab model and use a deterministic inference procedure, as in NExON-Bayes. Lastly, the third is the \textit{multiple graphical horseshoe} (mGHS) of \cite{Busatto02102025}, which treats networks exchangeably and uses a horseshoe prior on precision matrix entries along with an MCMC algorithm.

We principally assess performance based on \textit{precision}, which measures the proportion of predicted edges that are true positives, and \textit{recall}, which measures the proportion of the true edges that are correctly identified. For NExON-Bayes and covdepGE, we use a $0.5$ threshold on the edge PPI to determine edges, based on the median probability model rule \citep{barbieri2004optimal}. SSJGL does not estimate PPIs and so edges are determined to be present when the corresponding precision matrix entries are non-zero. For mGHS, we follow the $t^\alpha$ thresholding rule introduced by the authors \citep{Busatto02102025}, which uses a Metropolis-within-Gibbs step to update the threshold parameters. Figure \ref{fig:triple} shows the precision and recall of the estimates of each model for each network; for full results that include MCC, accuracy and sparsity, see Supplementary Material 2.3 Table S1. NExON-Bayes outperforms covdepGE with respect to precision, recall, MCC and accuracy. The poor performance of covdepGE in this setting can be explained by its overly sparse estimates. By only predicting a low number of edges, the total number of true positive edges will remain low, leading to very poor recall. The precision achieved is also the lowest of any model. Furthermore, covdepGE scales particularly poorly with the number of observations $N$, as it models the precision matrix as a continuous function of the covariate, leading to $|\mathcal{A}| \times N = 600$ precision matrix estimates rather than $|\mathcal{A}| = 4$ estimates returned by each of the other models. 

SSJGL and NExON-Bayes perform much more similarly in comparison, yet SSJGL estimates networks with slightly higher precision but much lower recall in most cases, particularly in networks $(1)$ and $(4)$. 
This is because SSJGL infers a shared similarity structure across all networks and uses it to inform the network estimates. In this simulated setting, edges that `appear' are mutual to networks $(2),(3) \text{ and } (4)$, whilst edges that `disappear' are mutual to networks $(1),(2) \text{ and } (3)$. Therefore, these edges are often estimated as present in all networks even if they are absent in networks $(1) \text{ and } (4)$ due to the model's tendency to over-estimate the level of similarity. 
This is explored further in Supplementary Material 2.2. 

\labelms{s1.10}mGHS exhibits the opposite problem to covdepGE, in that it estimates very dense networks, indicating poor convergence. It achieves the highest recall of any model, which is a misleading result common for such overly dense estimations. The precision of mGHS is the lowest of any model, leading to low accuracy compared to other models. mGHS performs far better in lower dimensional settings (see Supplementary Material 2.5), suggesting poor scalability.\labelms{e1.10} 

\begin{figure*}[h]
    \centering
    \includegraphics[width=\linewidth]{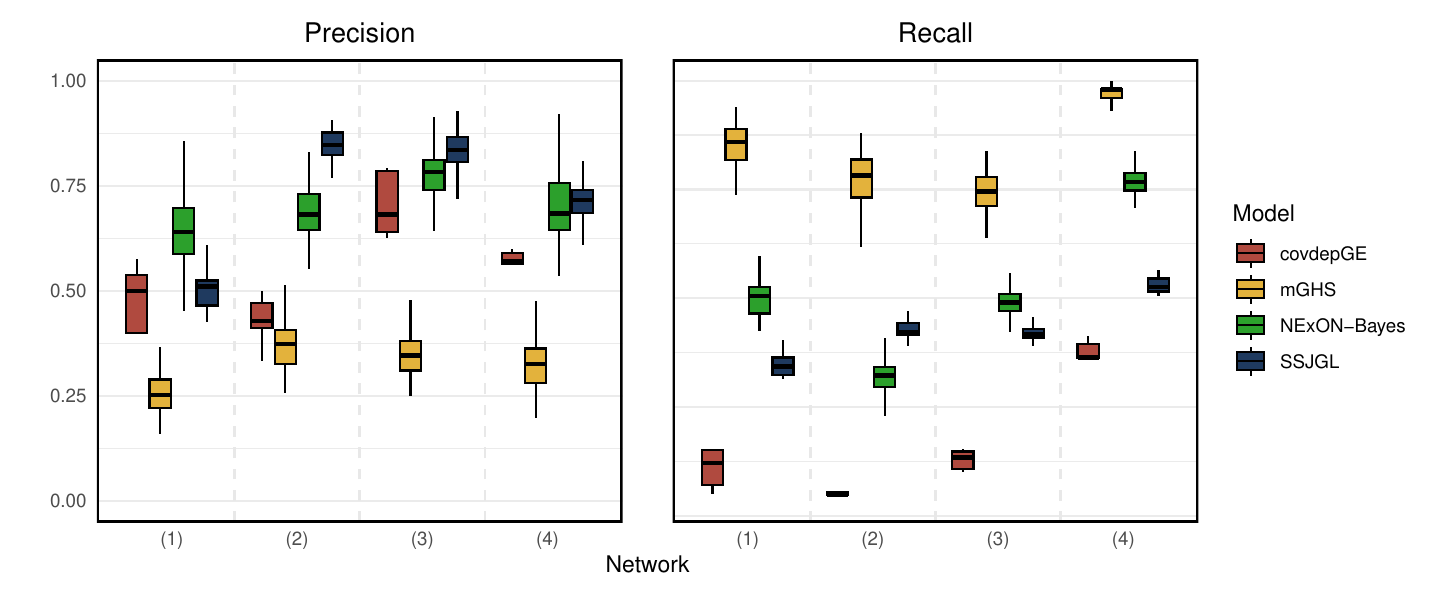}
    \caption{Performance of covdepGE, mGHS, NExON-Bayes \& SSJGL in the simulated scenario where there are $P = 100$ nodes and $N = 200$ samples for each of $\mid\mathcal{A} \mid = 4 $ networks. $50$ edges `appear' (absolute value of partial correlation linearly increases with covariate) and $50$ edges `disappear' (absolute value of partial correlation  linearly decreases with covariate). Performance is assessed over $100$ replicates for mGHS, NExON-Bayes and SSJGL and $10$ replicates for covdepGE due to excessive runtime for large $N$ ($20$+ hours).}
    \label{fig:triple}
\end{figure*}

\labelms{s1.5}Overall, this study exemplifies the advantages of NExON-Bayes in practice. To obtain a more complete assessment of NExON-Bayes' performance and test its robustness, three further simulation studies are explored that, to a varying degree, deviate from the specification of the model. Whilst the performance of NExON-Bayes worsens in misspecified frameworks (as expected), NExON-Bayes is shown to be robust in these settings by maintaining good performance metrics relative to its competitors. The full analysis of these additional simulation studies can be found in Supplementary Material 2.4.
\labelms{e1.5}

\labelms{s1.9}An analysis of computational runtimes indicates that NExON-Bayes performs inference more quickly than the other joint models that were tested (covdepGE, mGHS, SSJGL) within the range $10 \le P \le 250$. For up to 8 networks with up to 150 variables ($P \leq 150)$, NExON-Bayes runs in less than 30 minutes and in less than 2 seconds for $P \leq 50$, which is between 1 and 2 orders of magnitude quicker than its competitors. A more complete analysis is given in Supplementary Material 2.5.\labelms{e1.9}

\subsection{Real-World Application}\label{TCGA}
\begin{figure*}[]
    \centering
    \includegraphics[width=\linewidth]{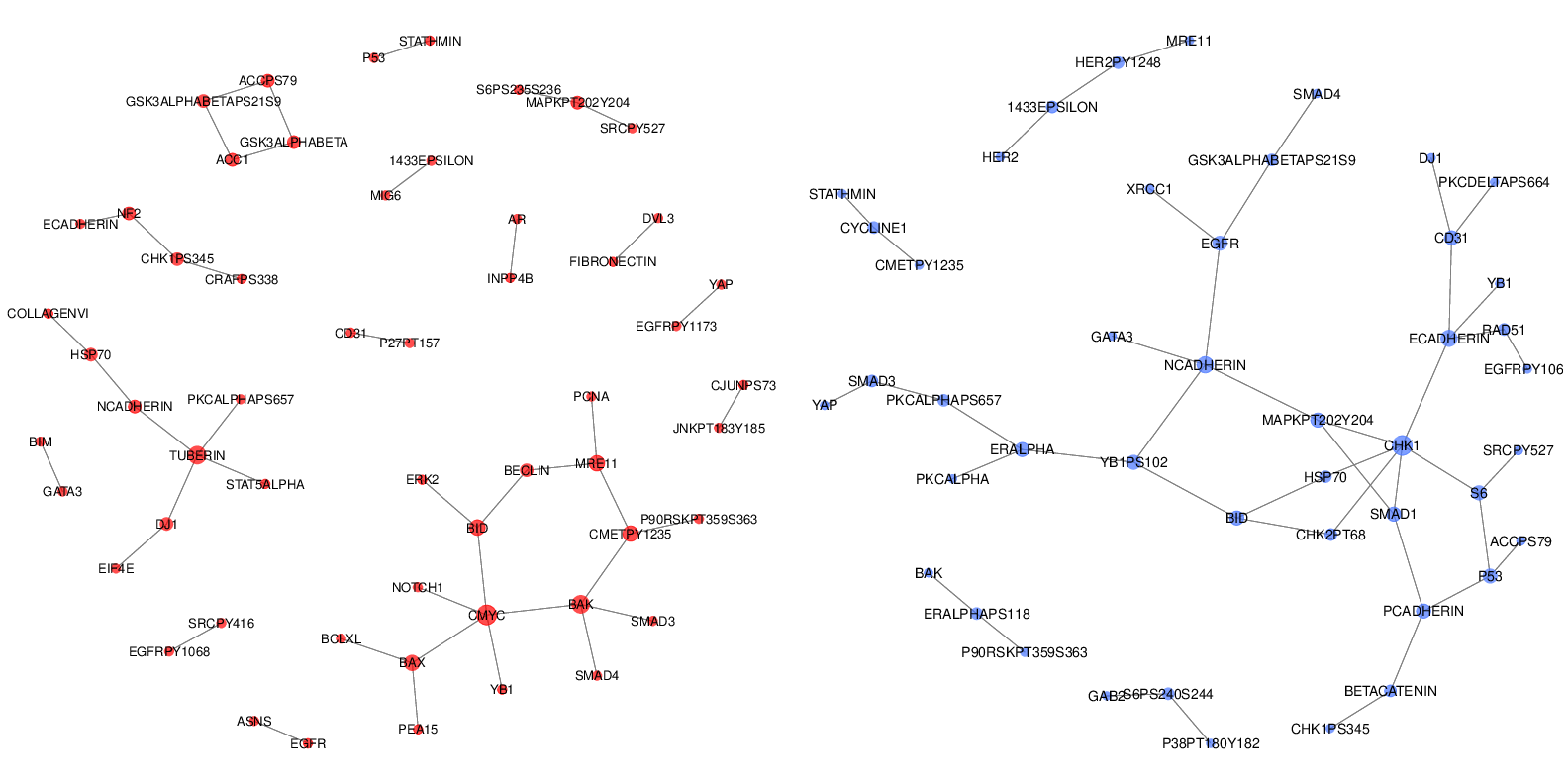}
    \caption{Subnetworks showing the edges with the largest absolute posterior point estimates of $\beta_{ij}$ in the BRCA application. The left network shows the edges with the largest positive $\beta_{ij}$ values, and the right network shows the edges with the 
    smallest negative $\beta_{ij}$ values.
}
    \label{fig:TCGA_betas}
\end{figure*}
We consider proteomic data from The Cancer Genome Atlas (TCGA) \citep{TCGA}, specifically, reverse phase protein array data of patients that have been diagnosed with stage I, II and III breast carcinoma\footnote{Stage I BRCA refers to tumours that have not spread beyond the breast and are still small in size, stage II refers to slightly larger tumours that may have spread to 1-3 lymph nodes, whilst stage III cancer refers to cancer that has spread to the `lymph nodes, the skin of the breast or the chest muscle' and is larger in size \citep{Mac}.} (BRCA). These data have been preprocessed using replicate-based normalisation \labelms{s1.6.2} \citep{goldman2020visualizing}, a commonly used method for omic data which eliminates unwanted batch effects whilst maintaining approximate Gaussianity, \labelms{e1.6.2}and contain the expression levels of $P = 131$ target proteins for $N_1 = 113$ stage I, $N_2=429$ stage II and $N_3= 140$ stage III individuals.
We treat tumour stage as our ordinal covariate capturing patient heterogeneity. Although factors like hormonal status also contribute to heterogeneity, we focus on tumour stage as an ordinal covariate to capture broad trends across disease progression. Our aim is to characterize how tumour severity impacts the proteomic network and whether conditional dependencies between protein pairs strengthen or weaken with tumour progression.

We firstly validate the estimations made by NExON-Bayes by consulting the STRING database \citep{szklarczyk2023string}, which contains protein-protein interactions that have been experimentally validated to a given confidence level. We calculate the percentage of pathways that are found by NExON-Bayes and SSL that have been validated to a 90\% confidence threshold and show that NExON-Bayes finds a higher percentage of these validated pathways than SSL for all three BRCA stages. \labelms{s1.7hub}Another useful analysis is to identify the `hub' proteins, which we define as proteins whose node degree exceeds the 90$^{th}$ percentile, in the networks estimated by NExON-Bayes. This can shed light on which proteins are prominent in which stages of BRCA. For example, NExON-Bayes identifies \textit{GATA binding protein 3}, \textit{Estrogen Receptor $\alpha$} and \textit{fibronectin} to have degrees in the top 5\% in all three stages, whilst \textit{Inositol polyphosphate-4-phosphatase (INPP4B)} is identified as a hub protein in only the stage III: \textit{INPP4B} inactivation is known to be common in triple-negative breast cancer cases \citep{Rodgers2021}; see Supplementary Material 3.1 \& 3.2 for full results of both of these analyses.\labelms{e1.7hub}

NExON-Bayes' framework provides unique insights into molecular networks by estimating the relation between the covariate and the conditional dependence. In essence, inspecting the edges corresponding to the largest absolute posterior point estimates of $\beta_{ij}$ could help identify biological processes that characterise tumour progression when used to estimate the proteomic networks of BRCA patients. Figure \ref{fig:TCGA_betas} shows the subnetworks consisting of the edges corresponding to the $50$ largest negative values (in absolute terms) and the $50$ largest positive values of the posterior mean estimate of $\beta_{ij}$.

To assess whether the borrowing of information facilitated through the inclusion of the edge-specific $\beta_{ij}$ is beneficial in this setting, we perform a ranked \textit{gene set enrichment analysis} (GSEA): \labelms{s1.7}genes are ranked by the sum of the estimated regression coefficients between the proteins which they encode and all other proteins (with positive and negative networks considered separately as in Figure \ref{fig:beta_mat}) and the resulting ranked list is then assessed to find whether genes belonging to predefined, biologically curated gene sets are over-represented towards the extremes (top or bottom) of the ranking, indicating non-random enrichment \citep{subramanian2005gene}.

We compare the gene sets found to be enriched in the $\beta_{ij}$ coefficients of NExON-Bayes (collectively denoted $\boldsymbol{\beta}_{\text{NExON}}$) with those found to be enriched in the coefficients estimated by GraphR ($\boldsymbol{\beta}_{\text{GraphR}}$), which is a graphical model that accounts for covariate information and adopts a decoupling strategy, wherein $\beta_{ij}$ coefficients are estimated separately before applying a variational Bayes inference algorithm. This differs to NExON-Bayes, which estimates each $\beta_{ij}$ and the precision matrix entries in conjunction within the VBECM algorithm. GraphR also enforces sparsity through a spike-and-slab prior on the precision matrices and uses  $\beta_{ij}$ to represent the effect of the covariate on edge-specific regression \citep{chen2025probabilistic}. This is very similar to NExON-Bayes and thereby provides a suitable and interesting comparison.
 GSEA finds 23 gene sets enriched in $\boldsymbol{\beta}_{\text{GraphR}}$, versus 22 in $\boldsymbol{\beta}_{\text{NExON}}$, where a 5\% threshold on the nominal $p$-value of each is employed, along with a 25\% false discovery rate threshold.
(Supplementary Material 3.3, Tables S5 \& S6).

The GSEA based on  $\boldsymbol{\beta}_{\text{GraphR}}$ identifies several sets related to \textit{epidermal growth factor receptor} (EGFR), which is a receptor tyrosine kinase often upregulated in various cancers, including breast cancer and colorectal cancer \citep{wee2017epidermal}.
Also identified are gene sets relating to \textit{ErbB protein signaling}. ErbB is a member of the EGFR family that is more prominently expressed in metastatic (stage IV) breast cancer than in primary (stage I) tumours \citep{xue2006erbb3}.
These sets are not found to be enriched in $\boldsymbol{\beta}_{\text{NExON}}$, although colorectal cancer-related pathways are, which could be explained by the confounding link to EGFR regulation.

Several of the gene sets found enriched in  $\boldsymbol{\beta}_{\text{NExON}}$ are either directly or indirectly related to apoptosis. The largest of these is the \textit{apoptosis modulation and signaling} pathway.\labelms{e1.7}
Apoptosis refers to programmed cell death, a vital process that maintains tissue homeostasis by eliminating damaged or unnecessary cells. In cancer, the evasion of apoptosis allows malignant cells to survive and proliferate uncontrollably. As such, pathways controlling apoptotic modulation and signaling play an important role in cancer progression \citep{plati2011apoptotic}.

Another enriched gene set of interest in this setting is the \textit{Tumor Necrosis Factor-$\alpha$ (TNF$\alpha$) signaling} pathway. TNF$\alpha$ is a proinflammatory cytokine that plays a significant role in various cellular processes, including apoptosis, proliferation, differentiation and immune responses. TNF$\alpha$ signaling can therefore either promote or inhibit tumor progression, depending on the cellular environment and specific signaling pathways activated \citep{10.1093/nar/gkad960}.

As a final example, the \textit{Ataxia Telangiectasia Mutated (ATM) Protein signaling} gene set is found enriched in the positive network of NExON-Bayes. The ATM signaling pathway plays a key role in breast cancer by maintaining genomic stability and influencing treatment responses. As an important regulator of the DNA damage response, ATM mutations can lead to increased susceptibility to malignancy and impact tumor progression \citep{VARADHAN2024155447}.

Whilst in this analysis there is no quantitative criterion  to ascertain which strategy (coupling vs decoupling) outperforms the other, these results show that both can find several relevant gene sets.

The ability to uncover these gene sets through encapsulating sample-level heterogeneity effectively in a joint model is an important practical advantage of NExON-Bayes (and GraphR). We have shown that, by taking this joint approach, interesting validated or candidate disease mechanisms are uncovered in a disease which progresses through defined stages. These mechanisms warrant dedicated investigation beyond the scope of this paper.
\labelms{e1.7}

\section{Discussion}\label{Discussion}
We have introduced a new method, NExON-Bayes, that jointly estimates multiple graphical models while accounting for sample-level heterogeneity as encapsulated in auxiliary ordinal covariate information. 
We show that, for such auxillary data, NExON-Bayes' formulation better captures sample-level heterogeneity than network estimation models that use continuous covariates or treat pre-defined groups of samples in an exchangeable manner, without resorting to covariate information. Moreover, thanks to its efficient deterministic inference procedure, our approach is more scalable and suited to the analysis of several current real data settings.
Applying our novel methodology to the TCGA breast cancer data set reveals new molecular pathways that characterize tumour stage progression, which could lead to new hypotheses about plausible oncogenic mechanisms.

Several avenues for future work could be considered. First, the current prior formulation assumes a linear relationship between the ordinal variable and the probability of edge inclusion -- relaxing this linearity assumption through an alternative monotonic parametric formulation may enhance both the model performance and its biological relevance.  
\labelms{s1.13}Secondly, the model could in principle be easily extended to accommodate multiple covariates. We did not pursue this extension here, however, as its practical usefulness is limited: indeed, it would require specifying a separate network for every combination of discrete covariate levels, which would reduce statistical power by fragmenting the sample across many networks. This would also hamper interpretability, as by combining multiple covariates to define networks, they lose their clear ordinality.\labelms{e1.13} 
\labelms{s1.8}Next, by using a deterministic inference algorithm, the model does not explicitly represent the full posterior distribution and may therefore underestimate posterior uncertainty \citep{giordano2018covariances}. In principle, stochastic inference algorithms can provide more faithful uncertainty quantification, provided convergence is reached. Accordingly, an interesting direction for future work is to investigate inference procedures that yield richer uncertainty quantification, for instance via hybrid deterministic–stochastic approaches, while maintaining a reasonable trade-off between computational cost and inferential accuracy.\labelms{e1.8}
\labelms{s2.7}
Another interesting extension concerns the treatment of missing data. 
Whilst NExON-Bayes does not intrinsically deal with missingness, several imputation methods are available for omics data \citep{lazar2016accounting, qiu2020genomic}. A natural extension to this work would be to instead incorporate model-based Bayesian imputation directly within NExON-Bayes by extending the model hierarchy to explicitly describe the observational process.\labelms{e2.7}
Lastly, as is typical for joint network estimation models, unequal sample sizes can lead to varying levels of sparsity in the estimated networks, reflecting differences in statistical power. This behaviour can in fact be desirable, as smaller sample sizes naturally yield sparser estimates, reducing the risk of false positives. However, if the goal is to isolate true biological variation independent of sample size effects, this pattern can be adjusted for — though doing so will require methodological advances beyond current approaches \citep{das2020nexus}.

NExON-Bayes offers a principled framework for joint network inference that incorporates sample-level covariates. By uncovering both shared and covariate-dependent patterns of conditional dependence, it enables the systematic characterization of sample-level heterogeneity and supports biologically grounded hypothesis generation for downstream validation.

\section*{Funding and access}
This research was supported by the Lopez–Loreta Foundation (J.F., H.R., C.L., X.X.) and the UK
Medical Research Council programme (J.F. and X.X.). C.L. has also received funding from the European Union’s Horizon Europe research and innovation programme under the Marie Sk\l odowska-Curie grant agreement No.101126636. For the purpose of open access, the authors have applied a Creative Commons Attribution (CC BY) license to any Author Accepted Manuscript version arising.

\bibliography{references}  
\bibliographystyle{plainnat}

\end{document}